\documentclass[12pt]{article}
\usepackage{a4wide,epsfig,amsmath,amssymb,cite,scalefnt,graphicx}
\numberwithin{equation}{section}
\allowdisplaybreaks
\usepackage{changepage}
\usepackage{tabularx}
\usepackage{graphics}
\usepackage{array}

\def\be{\begin{equation}}
\def\ee{\end{equation}}
\def\nn{\nonumber\\}

\def\Tr{\hbox {Tr}}

\def\psibar{\overline{\psi}}

\def\Tr{\mathop{\rm Tr}}
\def\det{\mathop{\rm det}}

\def\frakk[#1#2{{{#1}\over{#2}}}
\def\go{\rightarrow}

\def\Dhat{\hat D}
\def\Phat{\hat P}
\def\Qhat{\hat Q}
\def\phat{\hat p}
\def\qhat{\hat q}
\def\Pcal{\hat{\cal P}}
\def\Qcal{\hat{\cal Q}}
\def\Uhat{\hat U}
\def\Vhat{\hat V}
\def\onehat{\hat 1}

\def\psibar{\overline\psi}

\def\Vhat{\hat V}

\def\pa{\partial}

\input amssym.def
\input amssym
\def\pa{\partial}
\def\be{\begin{equation}}
\def\ee{\end{equation}}
\def\nn{\nonumber\\}

\def \nab{\nabla}

\def \R{{\cal R}}

\def \tsi{{\tilde \sigma}}

\def\Cbar{\bar{C}}

\def\cirk{\,{\raise1pt \hbox{${\scriptscriptstyle \circ}$}}\,}

\def \olr{{\raise6.5pt\hbox{$\leftrightarrow  \! \! \! \! \!$}}}

\def\Hcal{\cal H}
\def\Hcalhat{\hat\Hcal}
\def\hbar{\bar h}

\def\Rcal{{\cal R}}
\def\nabl{\overleftarrow \nabla}

\font\ninerm=cmr9 \font\ninesy=cmsy9
\font\eightrm=cmr8 \font\sixrm=cmr6
\font\eighti=cmmi8 \font\sixi=cmmi6
\font\eightsy=cmsy8 \font\sixsy=cmsy6
\font\eightbf=cmbx8 \font\sixbf=cmbx6
\font\eightit=cmti8
\def\eightpoint{\def\rm{\fam0\eightrm}
  \textfont0=\eightrm \scriptfont0=\sixrm \scriptscriptfont0=\fiverm
  \textfont1=\eighti  \scriptfont1=\sixi  \scriptscriptfont1=\fivei
  \textfont2=\eightsy \scriptfont2=\sixsy \scriptscriptfont2=\fivesy
  \textfont3=\tenex   \scriptfont3=\tenex \scriptscriptfont3=\tenex
  \textfont\itfam=\eightit  \def\it{\fam\itfam\eightit}%
  \textfont\bffam=\eightbf  \scriptfont\bffam=\sixbf
  \scriptscriptfont\bffam=\fivebf  \def\bf{\fam\bffam\eightbf}%
  \normalbaselineskip=9pt
  \setbox\strutbox=\hbox{\vrule height7pt depth2pt width0pt}%
  \let\big=\eightbig  \normalbaselines\rm}
\catcode`@=11 %
\def\eightbig#1{{\hbox{$\textfont0=\ninerm\textfont2=\ninesy
  \left#1\vbox to6.5pt{}\right.\n@@space$}}}
\def\vfootnote#1{\insert\footins\bgroup\eightpoint
  \interlinepenalty=\interfootnotelinepenalty
  \splittopskip=\ht\strutbox %
  \splitmaxdepth=\dp\strutbox %
  \leftskip=0pt \rightskip=0pt \spaceskip=0pt \xspaceskip=0pt
  \textindent{#1}\footstrut\futurelet\next\fo@t}
\catcode`@=12 %
\def\today{\number\day\ \ifcase\month\or January\or February\or March\or
April\or May\or June\or July\or
August\or September\or October\or November\or December\fi, \number\year}

\input epsf

\begin{document}

\begin{titlepage}
\begin{flushright}
LTH1229
\end{flushright}
\date{}
\vspace*{3mm}

\begin{center}
{\Huge One-loop $\beta$-functions for renormalisable gravity}\\[12mm]
{\bf I.~Jack}\\

\vspace{5mm}
Dept. of Mathematical Sciences,
University of Liverpool, Liverpool L69 3BX, UK\\

\end{center}

\vspace{3mm}

\begin{abstract}
We compute the one-loop $\beta$-functions for renormalisable quantum gravity coupled to scalars using the co-ordinate space approach and generalised Schwinger De Witt technique. We resolve apparent contradictions with the corresponding momentum space calculations, and indicate how our results also resolve similar inconsistencies in the fermion case. 
\end{abstract}

\vfill

\end{titlepage}
\section{Introduction}
Higher-derivative quantum gravity has attracted attention over the years as a renormalisable alternative to Einstein gravity; hence the alternative name of renormalisable quantum gravity. It is characterised by the addition to the action of terms quadratic in the Riemann curvature tensor. Perturbative calculations are involved and technically challenging, and even in the pure gravity case it took considerable effort to obtain the correct one-loop $\beta$-functions. The first computation in Ref.~\cite{JT} was corrected in Ref.~\cite{FT}; but the final correct result was obtained in Refs.~\cite{Av2}, \cite{Av3} using the generalised Schwinger-De Witt technique presented in Ref.~\cite{BV}. A comprehensive overview of the calculation may be found in Ref.~\cite{Av1} where some typos present in the Riemann tensor expansions in Ref.~\cite{Av2} are corrected. The case of higher-derivative quantum gravity coupled to scalars and fermions was considered in Ref.~\cite{BOS}, and results were presented for the one-loop $\beta$ functions. In recent years, the subject of higher-derivative quantum gravity coupled to matter has attracted renewed interest in the context of ``Agravity''\cite{SS}, where the Planck scale arises dynamically at quantum level. Independently in Ref.~\cite{EJ0} the possibility of dimensional transmutation within higher-derivative quantum gravity leading to an effective Einstein-Hilbert theory was raised and further explored in Refs.~\cite{EJ0a}-\cite{EJ4}. In Ref.~\cite{SS} the gravitational $\beta$-functions were rederived using momentum space techniques. It was pointed out in Ref.~\cite{EJ1} that these $\beta$-functions disagreed significantly with those derived in Ref.~\cite{BOS}, and simple arguments were presented to show that the results in Ref.~\cite{BOS} could not be correct. This is clearly an unsatisfactory state of affairs. Co-ordinate space methods have the advantage for calculations in gravity and curved spacetime of being manifestly generally co-ordinate invariant; and the generalised Schwinger-De Witt technique in particular is a powerful and elegant technique for computations in higher-derivative theories. It would be comforting to be assured that there are no issues of principle involved and that the technique will reliably reproduce momentum space results; both for future calculations and in existing cases where there are no momentum space results for comparison. Accordingly we have carefully reproduced the one-loop $\beta$-function computations of Refs.~\cite{BOS} for the case of higher-derivative gravity coupled to scalars, and, for completeness, the pure gravity case considered in  Ref.~\cite{Av1}. We agree precisely with the results of Ref.~\cite{Av1} though, as mentioned, some of the intermediate equations contain typos. The calculations of Ref.~\cite{BOS} are mostly correct after once again allowing for some typos; however, we have identified some discrepancies after a close scrutiny of the application of the Schwinger-De Witt technique, which conceals some subtleties in the compact notation of Ref.~\cite{BV}. We present our version of the calculation here. It seems appropriate to provide plenty of detail, both in order to give definitive versions of all the results and in order to carefully explain the origin of the new terms we have identified; though where possible we have relegated the details to Appendices.
\section{Outline of calculation and results}
We work with a Lorentz signature $(+---)$  and a Riemann tensor defined by 
\be
R^{\mu}{}_{\nu\alpha\beta}=\pa_{\alpha}\Gamma^{\mu}{}_{\nu\beta}-\ldots
\ee
so that 
\be
[\nabla_{\alpha},\nabla_{\beta}]v^{\mu}=R^{\mu}{}_{\nu\alpha\beta}v^{\nu}.
\ee
The classical action is given by
\be
S=\int dv_x\Bigl\{\frac{1}{2\mu}C^{\mu\nu\rho\sigma}C_{\mu\nu\rho\sigma}-\frac{\omega}{3\mu}R^2+\tfrac12g^{\mu\nu}\pa_{\mu}\phi\pa_{\nu}\phi+\tfrac12\xi R\phi^2-\tfrac{1}{24}\lambda\phi^4\Bigr\},
\label{action}
\ee
where $dv_x=d^4x\sqrt{-g}$ and the Weyl tensor $C_{\mu\nu\rho\sigma}$ is defined by
\be
C_{\mu\nu\rho\sigma}=R_{\mu\nu\rho\sigma}+\tfrac12(R_{\mu\sigma}g_{\nu\rho}-R_{\mu\rho}g_{\nu\sigma}
+R_{\nu\rho}g_{\mu\sigma}-R_{\nu\sigma}g_{\mu\rho})+\tfrac16R(g_{\mu\rho}g_{\nu\sigma}-g_{\mu\sigma}g_{\nu\rho}),
\ee
so that
\be
C^{\mu\nu\rho\sigma}C_{\mu\nu\rho\sigma}=R^{\mu\nu\rho\sigma}R_{\mu\nu\rho\sigma}-2R^{\mu\nu}R_{\mu\nu}+\frac13R^2
=2\left(R^{\mu\nu}R_{\mu\nu}-\frac13R^2\right),
\ee
where we have imposed the Gauss-Bonnet identity $R^{\mu\nu\rho\sigma}R_{\mu\nu\rho\sigma}-4R^{\mu\nu}R_{\mu\nu}+R^2=0$ in $d=4$. The gravitational terms in Eq.~\eqref{action} may then be rewritten as $\frac{1}{\mu}\left[R^{\mu\nu}R_{\mu\nu}-\frac13(1+\omega)R^2\right]$.
We now expand around a classical background metric and scalar field according to 
\be
g_{\mu\nu}\rightarrow g^q_{\mu\nu}=g_{\mu\nu}+h_{\mu\nu},\quad 
\phi\rightarrow \phi^q=\phi-i\tsi,
\ee
(using for convenience the same decomposition as in Ref.~\cite{BOS}) and define 
\be
h^*=h^{\alpha}{}_{\alpha},\quad \hbar_{\mu\nu}=\frac{1}{\sqrt{2\mu}}(h_{\mu\nu}-\tfrac14g_{\mu\nu}h^*),\quad h=\sqrt{-\frac{\beta}{2\mu}}h^*,
\ee
where $\hbar_{\mu\nu}$ is tracefree.
The effective action is given by
\begin{align}
Z[g_{\mu\nu},\phi]=&(\det G^{\mu\nu})^{-\tfrac12}\int D\hbar_{\mu\nu}DhD\Cbar_{\alpha}DC^{\beta}\exp\Bigl\{i\Bigl[S(g^q,\phi^q)-\frac{\delta S}{\delta g_{\mu\nu}}h_{\mu\nu}\nn&-\frac{\delta S}{\delta \phi}(-i\tsi)+\tfrac12\chi_{\mu}G^{\mu\nu}\chi_{\nu}
+\Cbar_{\alpha}M^{\alpha}{}_{\beta}C^{\beta}\Bigr]\Bigr\},
\label{Zdef}
\end{align}
where $S$ is defined in Eq.~\eqref{action} and where
\be
M^{\alpha}{}_{\beta}=G^{\alpha\mu}\frac{\delta \chi_{\mu}}{\delta g_{\rho\sigma}}R_{\rho\sigma,\beta}
\ee
with 
\be
R_{\rho\sigma,\beta}=g_{\rho\beta}\nab_{\sigma}+g_{\sigma\beta}\nab_{\rho}
\ee
and where we take the gauge $\chi_{\mu}$ and the weight functional to be given by
\begin{subequations}\label{gauge1:main}
\begin{align}
\chi_{\mu}=&\nab_{\alpha}h^{\alpha}{}_{\mu}-(\beta+\tfrac14)\nab_{\mu}(h^{\alpha\beta}g_{\alpha\beta}),\\
G^{\mu\nu}=&\frac{1}{\alpha}(-g^{\mu\nu}\Box-\gamma\nab^ {\mu}\nab^{\nu}+\nab^{\nu}\nab^{\mu}+{\cal P}^{\mu\nu}).
\end{align}
\end{subequations}
We also take the gauge parameters $\alpha$, $\beta$ and $\gamma$ to be given by
\be
\alpha=1,\quad \beta=\frac{3\omega}{4(1+\omega)},\quad \gamma=\frac23(1+\omega),
\label{gauge2}
\ee
which ensures a minimal form for the four-derivative terms in the expansion; and we take ${\cal P}^{\mu\nu}=0$.

The quadratic terms in the exponent in Eq.~\eqref{Zdef} now take the form
\be
\tfrac12i\int dv_x\left(\begin{matrix}\hbar_{\rho\sigma}&h&\tsi\end{matrix}\right)
\Hcalhat \left(\begin{matrix}\hbar_{\mu\nu}\cr h\cr \tsi\end{matrix}\right),
\ee
where the matrix operator $\Hcalhat$ is given by
\be
\Hcalhat=\left(\begin{matrix}\onehat\Box^2+\Vhat+\Uhat&\qhat_1+\qhat_2+\qhat_3\cr
\phat_1+\phat_2+\phat_3&
\Box+\Dhat
\label{Hdef}
\end{matrix}\right),
\ee
where
\be
\Vhat=\Vhat^{\alpha\beta}\nab_{\alpha}\nab_{\beta}
\ee
and
\begin{subequations}\label{Pform:main}
\begin{align}
\phat_1=&\tfrac12(\phat_1^{\alpha\beta}\nab_{\alpha}\nab_{\beta}+\nabl_{\alpha}\nabl_{\beta}\qhat_1^{T\alpha\beta}),\label{Pform:a}\\
\qhat_1=&\tfrac12(\qhat_1^{\alpha\beta}\nab_{\alpha}\nab_{\beta}+\nabl_{\alpha}\nabl_{\beta}\phat_1^{T\alpha\beta}),\\
\phat_2=&\tfrac12(\phat_2^{\alpha}\nab_{\alpha}+\nabl_{\alpha}\qhat_2^{T\alpha}),\\
\qhat_2=&\tfrac12(\qhat_2^{\alpha}\nab_{\alpha}+\nabl_{\alpha}\phat_2^{T\alpha}).
\end{align}
\end{subequations}
Here we see the first major difference between our calculation and that of Ref.~\cite{BOS}. We believe that the form of Eq.~\eqref{Pform:main} encodes the requisite symmetry properties of the matrix operator $\Hcalhat$, for arbitrary $\phat_1$,
$\qhat_1$, $\phat_2$, $\qhat_2$; though the quadratic expansion naturally (i.e. without any integration by parts) results in 
\be 
\qhat_1^{\alpha\beta}=\phat_2^{\alpha}=0.
\label{chjj}
\ee
We denote the ``natural'' (in the above sense) forms of the other quantities in  Eq.~\eqref{Pform:main}, together  with $\phat_3$, $\qhat_3$, by 
\be
\phat_1^{\alpha\beta}=\Pcal_1^{\alpha\beta},\quad \qhat_2^{\alpha}=\Qcal_2^{\alpha},\quad \phat_3=\Pcal_3,\quad \qhat_3=\Qcal_3.
\label{chjja}
\ee
Explicit expressions for $\Pcal_1^{\alpha\beta}$, $\Qcal_2^{\alpha}$, etc, will be given later, in Appendix A. 

In Ref.~\cite{BOS}, on the other hand, the quadratic expansion is rewritten using integration by parts to obtain $\phat^{\alpha\beta}_1$,
$\qhat^{\alpha\beta}_1$, $\phat^{\alpha}_2$, $\qhat^{\alpha}_2$, satisfying 
\begin{subequations}\label{Pforma:main}
\begin{align}
\phat_1^{\alpha\beta}=&\qhat_1^{T\alpha\beta}=\Phat_1^{\alpha\beta},\label{Pforma:a}\\
\qhat_2^{\alpha}=&-\phat_2^{T\alpha}=\Qhat_2^{\alpha},
\end{align}
\end{subequations} 
and the  corresponding forms of $\phat_3$, $\qhat_3$ are similarly denoted
\be
\phat_3=\Phat_3,\quad \qhat_3=\Qhat_3.
\ee
Again, the explicit expressions for $\Phat_1^{\alpha\beta}$, $\Qhat_2^{\alpha}$ etc are postponed to Appendix A.

However, the terms with left-acting and right-acting derivatives are then (mistakenly, we believe), conflated in Ref.~\cite{BOS}, so that, for instance, $\phat_1=\tfrac12(\Phat_1^{\alpha\beta}\nab_{\alpha}\nab_{\beta}+\nabl_{\alpha}\nabl_{\beta}\Phat_1^{\alpha\beta})$ as follows from Eqs.~\eqref{Pform:a}, \eqref{Pforma:a} is replaced  by 
$\phat_1=\Phat_1^{\alpha\beta}\nab_{\alpha}\nab_{\beta}$.
On the other hand, we believe that the symmetrisation in Eq.~\eqref{Pforma:main} is an unnecessary complication, and we show in the appendices that the correct answer is obtained either if Eq.~\eqref{Pforma:main} is imposed or in the simpler case (Eqs.~\eqref{chjj}, \eqref{chjja} where it isn't; provided that the left-acting derivatives in Eq.~\eqref{Pform:main} are retained. Note that the quantities appearing on the diagonal in Eq.~\eqref{Hdef}, namely $\Uhat$, $\Vhat$ and $\Dhat$, are unaffected by this process of integrating by parts and have the same form whichever choice is made.  We also emphasise the slight change of notation relative to Ref.~\cite{BOS}, in that we have used lower-case letters to denote general quantities such as $\phat_1^{\alpha\beta}$, $\qhat_1^{\alpha\beta}$, reserving upper-case letters $\Phat^{\alpha\beta}_1$, $\Qhat^{\alpha\beta}_1$ to denote the particular forms assigned to these quantities in Ref.~\cite{BOS}, and using $\Pcal^{\alpha\beta}_1$, $\Qcal^{\alpha\beta}_1$ to denote the corresponding forms of these quantities in our simpler version of the calculation. Our corrected version of the computation in Ref.~\cite{BOS} will be described in more detail in Appendices A and B, while our simpler version is contained in Appendices A and C.

The effective action corresponding to Eq.~\eqref{Zdef} is now given up to one loop by 
\be
i\Gamma=iS+\frac12\Tr\ln\Hcalhat-\Tr\ln M^{\alpha}{}_{\beta}-\tfrac12\Tr\ln G_{\mu\nu},
\ee
with $S$ as in Eq.~\eqref{action} and where the functional trace $\Tr$ is defined by
\be
\Tr A=\int dv_xA(x,y)|_{y=x}.
\ee
We now rewrite $\Tr\ln\Hcalhat$ in the convenient form
\be
\Tr\ln\Hcalhat=\Tr\ln\left(\begin{matrix}\onehat\Box^2&0\\
0&\Box
\end{matrix}\right)
+\Tr\ln\Bigl[\left(\begin{matrix}\onehat&0\\
0&\Box
\end{matrix}
\right)
+\Bigl(\begin{matrix}V+U&q_1+q_2+q_3\\
p_1+p_2+p_3&D
\end{matrix}\Bigr)\Bigr],
\label{trace}
\ee
where
\begin{subequations}
\begin{align}
\onehat=&\left(\begin{matrix} \delta^{\mu\nu,\rho\sigma}-\tfrac14g^{\mu\nu}g^{\rho\sigma}&0\cr
                                                             0&1\end{matrix}\right),\\
\delta^{\mu\nu,\rho\sigma}=&\tfrac12(g^{\mu\rho}g^{\nu\sigma}+g^{\mu\sigma}g^{\nu\rho}),
\end{align}
\end{subequations}
and
\begin{align}
V=\Vhat\frac{\onehat}{\Box^2},\quad U=&\Uhat\frac{\onehat}{\Box^2},\quad D=\Dhat\frac{1}{\Box}\nn
p_1=\phat_1\frac{\onehat}{\Box^2}, \quad p_2=&\phat_2\frac{\onehat}{\Box^2},\quad p_3=\phat_3\frac{\onehat}{\Box^2}, \nn
q_1=\qhat_1\frac{1}{\Box},\quad q_2=&\qhat_2\frac{1}{\Box},\quad q_3=\qhat_3\frac{1}{\Box}. 
\label{Pbox}
\end{align}
Note the appearance of $\frac{\onehat}{\Box^2}$ in Eq.~\eqref{Pbox}; this is a shorthand notation for the inverse of the operator $\Box^2 \onehat$. We can now rewrite the second term in Eq.~\eqref{trace} in the form
\begin{align}
\Tr\ln\Hcalhat=&\Tr\ln(\Box+\Dhat)\nn
&+\Tr\ln(-p_1q_1-p_1q_2-p_2q_1-p_2q_2\nn
&-p_1q_3-p_3q_1+Vq_1p_1+q_1Dp_1\nn
&-\tfrac12q_1p_1q_1p_1+U+V+D-\tfrac12V^2-\tfrac12D^2)+\ldots
\label{exp}
\end{align}
The strategy now is to commute the operator $\frac{1}{\Box}$ through the other quantities until all the Green functions are at the right-hand side, using
\be
\frac{1}{\Box}X=X\frac{1}{\Box}+\frac{1}{\Box}[X,\Box]\frac{1}{\Box}.
\label{comm}
\ee
Derivatives will then act either on functions of $x$ such as $\phat_3$, $\qhat_3$, $\Vhat^{\alpha\beta}$, $\Uhat$ or $\Dhat$ in Eq.~\eqref{Hdef}, or $\phat_1^{\alpha\beta}$, $\qhat_1^{\alpha\beta}$,
$\phat_2^{\alpha}$, $\qhat_2^{\alpha}$ in Eq.~\eqref{Pform:main}; or on products of $\frac{1}{\Box}$. The latter are the ``universal functional traces'' of Ref.~\cite{BV}, where they are listed and their divergences given (using dimensional regularisation with dimension $d=4-\epsilon$). Eq.~\eqref{nabid} may then be used to extract the divergences. The details are given
in the appendices. Here we just remark that another major source of disagreement with Ref.~\cite{BOS} is that when $X$ is a quantity such as $\phat^{\alpha\beta}_1\nab_{\alpha}\nab_{\beta}$,
it is important to include terms resulting from commuting $\Box$ with the derivatives $\nab_{\alpha,\beta}$. Furthermore, in doing this it is important to recall the tensor nature of $\frac{\onehat}{\Box^2}$, so that
\be
[\nab_{\alpha},\nab_{\beta}]\onehat=\R_{\alpha\beta}\onehat,
\ee
where $\R_{\alpha\beta}$ is defined in Eq.~\eqref{Rcal}. We must also include the terms involving $\R$ in Eq.~\eqref{nabid} where appropriate; these were overlooked in Ref.~\cite{BOS} .

In Appendix A we give expressions for the quantities in Eqs.~\eqref{Hdef} together with results forthe divergent contributions from  individual terms in the expansion Eq.~\eqref{exp}. We also list the relevant results for divergent parts of the universal functional traces. However, the terms in the expansion Eq.~\eqref{exp} which depend on whether the choice Eq.~\eqref{chjj}
or Eq.~\eqref{Pforma:main} is made are considered separately in Appendices B, C respectively. The sums of the divergent terms, of course, are independent of the choice made. Hence, adding either Eqs.~\eqref{noPQ:x}-\eqref{noPQ:y} and Eqs.~\eqref{BOSres:x}-\eqref{BOSres:y}, or Eqs.~\eqref{noPQ:x}-\eqref{noPQ:y} and Eqs.~\eqref{jjres:x}-\eqref{jjres:y}, we obtain the one-loop divergences, in the form
\begin{align}
\left(\tfrac12\Tr\ln\Hcalhat\right)_{\rm{div}}=&i\int dv_x\Bigl[\tfrac12\zeta_{\phi}g^{\mu\nu}\pa_{\mu}\phi\pa_{\nu}\phi+\tfrac12\zeta_{\xi } R\phi^2-\tfrac{1}{24}\zeta_{\lambda}\phi^4+\ldots\Bigr],
\end{align}
where we omit the purely curvature-dependent divergences, and where
\begin{subequations}
\begin{align}
\zeta_{\phi}=&\frac{\mu}{16\pi^2\epsilon}\left\{-3\xi\left(1+\frac{1}{4\beta}\right)+\frac94-\frac{1}{16\beta}\right\},\\
\zeta_{\xi}=&\frac{1}{16\pi^2\epsilon}\Bigl[-\left(\xi-\tfrac16\right)\lambda+\mu\xi\Bigl\{\xi^2\left(-3+\frac{9}{4\beta}\right)+\xi\left(\frac12+\omega-\frac{15}{8\beta}-\frac{9\omega}{16\beta^2}\right)\nn
&+\left(\frac74-\frac72\omega+\frac{1}{16\beta}+\frac{3\omega}{32\beta^2}\right)\Bigr\}\Bigr],\\
\zeta_{\lambda}=&\zeta_{\lambda_1}\lambda+\zeta_{\lambda_2},\nn
\zeta_{\lambda_1}=&\frac{1}{16\pi^2\epsilon}\Bigl[-3\lambda+\mu\Bigl\{\xi^2\left(-18+\frac{27}{2\beta}\right)-\frac{6\xi}{\beta}+\left(9+\frac{1}{4\beta}\right)\Bigr\}\Bigr],\\
\zeta_{\lambda_2}=&\frac{\mu^2}{16\pi^2\epsilon}6\xi^2\Bigl\{
\xi^2\left(-\frac92+\frac{27}{4\beta}-\frac{81}{32\beta^2}\right)
+\xi\left(\frac32-\frac{9}{4\beta}+\frac{27}{32\beta^2}\right)\nn
&+\left(-\frac{21}{8}+\frac{3}{16\beta}-\frac{9}{128\beta^2}\right)\Bigl\}
\end{align}
\end{subequations}
The corresponding results for individual $\zeta_{\phi}$, $\zeta_{\lambda}$ and $\zeta_{\xi}$ are not given in Ref.~\cite{BOS}; but it is worth pointing out here that one significant difference between our results and those of Ref.~\cite{BOS} is that their result for $\zeta_{\phi}$, in contrast to ours, would contain terms proportional to $\xi^2$. 
After making the replacements
\be
\phi\go Z_{\phi}^{\tfrac12}\phi,\quad \xi\go \xi_B, \quad \lambda\go \lambda_B,
\label{reg}
\ee
we find that finiteness at this order may be ensured by taking
\begin{align}
Z_{\phi}^{(1)}=&-\zeta_{\phi},\nn
\xi_B^{(1)}=&\zeta_{\phi}\xi-\zeta_{\xi}=\frac{1}{16\pi^2\epsilon}\Bigl[\left(\xi-\tfrac16\right)\lambda+\mu\xi\Bigl\{\xi^2\left(3-\frac{9}{4\beta}\right)+\xi\left(-\frac72-\omega+\frac{9}{8\beta}+\frac{9\omega}{16\beta^2}\right)\nn
&+\left(\frac12+\frac72\omega-\frac{1}{8\beta}-\frac{3\omega}{32\beta^2}\right)\Bigr\}\Bigr],\\
\lambda_B^{(1)}=&2\zeta_{\phi}\lambda-\zeta_{\lambda}=\frac{1}{16\pi^2\epsilon}\Bigl[3\lambda^2\nn
&+\mu\lambda\Bigl\{\xi^2\left(18-\frac{27}{2\beta}\right)+\xi\left(-6+\frac{9}{2\beta}\right)-\left(\frac92+\frac{3}{8\beta}\right)\Bigr\}\\
&+6\mu^2\xi^2\Bigl\{
\xi^2\left(\frac92-\frac{27}{4\beta}+\frac{81}{32\beta^2}\right)
+\xi\left(-\frac32+\frac{9}{4\beta}-\frac{27}{32\beta^2}\right)\nn
&+\left(\frac{21}{8}-\frac{3}{16\beta}+\frac{9}{128\beta^2}\right)\Bigl\}\Bigr].
\end{align}
We emphasise for later reference that the extra $\xi^2$ terms included in $\zeta_{\phi}$ in Ref.~\cite{BOS} will induce extra $\xi^2$ terms in $Z_{\phi}^{(1)}$, and also in $\xi_B^{(1)}$ and $\lambda_B^{(1)}$.
Of course this would have effects on their results for the $\beta$-functions for $\xi$ and $\lambda$.
The one-loop $\beta$-functions are then given as usual by
\be
\beta^{(1)}_\xi=\epsilon  \xi_B^{(1)},
\ee
etc. After substituting for the gauge parameter $\beta$ from Eq.~\eqref{gauge2}, we obtain
\begin{subequations}
\begin{align}
16\pi^2\beta_\lambda=&3\lambda^2-\mu\lambda\left\{\frac{18}{\omega}\left(\xi-\frac16\right)^2+5\right\}\\
&+6\mu^2\xi^2\left\{\frac{9}{2\omega^2}\left(\xi-\frac16\right)^2+\frac52\right\},\nn 
16\pi^2\beta_\xi=&\left(\xi-\tfrac16\right)\lambda+\mu\xi\left\{\frac{10}{3}\omega-\frac{1}{\omega}\left(\xi-\frac16\right)\left(3\xi-2\right)\right\},
\end{align}
\end{subequations}
These results now agree with the momentum-space calculations of Ref.~\cite{SS} used in Ref.\cite{EJ1}. To facilitate comparison, the relation between our parameters and those used in Refs.~\cite{SS}, \cite{EJ1} is as follows:
\be 
\mu=-f_2^2=-a,\quad\omega=\frac{f_2^2}{2f_0^2}=-\frac{a}{b},
\ee
while our $\xi$ has the opposite sign to that in Refs.~\cite{SS}, \cite{EJ1}.

The $\beta$-function for a Yukawa coupling in renormalisable gravity was presented in Ref.~\cite{SS} with the comment that no earlier result could be found in the literature. In fact the position space calculation  of this $\beta$-function was performed in Ref.~\cite{BOS}, for the case of theories with three-dimensional multiplets of scalar and fermion fields. It is simple to adapt their discussion to a case more akin to that considered in Ref.~\cite{SS}, where we add to Eq.~\eqref{action} the action for a single Weyl fermion
\be
S_F=\int dv_x[i\psibar\gamma^{\mu}\pa_{\mu}\psi-(y\phi\psi\psi+\hbox{h.c.})].
\label{fermact}
\ee
The theory is regularised by Eq.~\eqref{reg} together with
\be
y\go y_B, \quad \psi\go Z_{\psi}^{\tfrac12}\psi,
\ee
and the one-loop $\beta$-function is given as usual by
\be
\beta_y^{(1)}=\epsilon y_B^{(1)}.
\ee
The results analogous to $y_B^{(1)}$, $Z^{(1)}_{\psi}$ are not given separately in Ref.~\cite{BOS}. However, the final result quoted for  the one-loop Yukawa $\beta$-function clearly differs from the corresponding result in Ref.~\cite{SS} in having two terms proportional to $\xi^2$. In fact, a closer scrutiny reveals that these terms precisely correspond to the $\xi^2$-dependent terms erroneously included in the result of Ref.~\cite{BOS} for $Z_{\phi}^{(1)}$, as remarked earlier. Upon removing these terms, the result of Ref.~\cite{BOS} when adapted to the fermion action in Eq.~\eqref{fermact} becomes
\be
16\pi^2\beta_y^{(1)}=\frac92y^3-\frac{15}{8}\mu y
\ee
and the results of Refs.~\cite{SS} and \cite{BOS} then agree precisely. We have not repeated the remainder of the position space computation by calculating $Z^{(1)}_y$, $Z^{(1)}_{\psi}$ independently, but this agreement is strong evidence in favour of its accuracy.

\section{Conclusions}
We have resolved the disagreements between the position space and momentum space calculations of the one-loop $\beta$-functions for renormalisable (i.e. higher derivative) gravity coupled to scalar and fermion fields. The balance of probability was always in favour of the momentum space result presented in Ref.~\cite{SS}, since it had been computed in two independent ways (albeit within the same paper) and since it could be argued\cite{EJ1} that the result of Ref.~\cite{BOS} could not be correct. Nevertheless in the face of two competing results it is reassuring to have a completely independent confirmation of one of them. Furthermore, it is also comforting to know that there are no ineradicable issues of principle in the position space computation, in view of its elegance and its manifest covariance. It is worth pointing out that the position space calculation can be carried out by hand at this loop level, whereas computer packages were enlisted in the momentum space case. 

An interesting feature of the $\beta$-functions for the couplings $\lambda$ and $\xi$ is the considerable simplification that takes place when $\xi$ takes the classically conformal value $\xi=\tfrac16$. This is by no means guaranteed {\it a priori} even in the case of a curved background, still less in the case of quantum gravity. It would be interesting to investigate whether such behaviour persists to higher orders; or in other dimensions such as $d=6$ where the corresponding conformal value is $\xi=\tfrac15$. Much of the position space calculation for $d=6$ could easily be  adapted from the current calculation for $d=4$.

We have used the same choice of gauge as in Ref.~\cite{BOS}, namely taking the values listed in Eq.~\eqref{gauge2} in Eq.~\eqref{gauge1:main}, and taking a minimal choice ${\cal P}^{\mu\nu}=0$. These choices for $\alpha$, $\beta$, $\gamma$ are required by the generalised Schwinger-De Witt technique, by guaranteeing a minimal form for the four-derivative operator in Eq.~\eqref{Hdef}. However, in the pure gravity case, the independence at least of the choice of ${\cal P}^{\mu\nu}$ was demonstrated in Ref.~\cite{Av1}. It is less clear how to proceed to show this independence in the current case with couplings to scalar and fermion fields. It will be important to resolve this issue before tackling the $d=6$ calculation, where it is not {\it a priori} clear what is the minimal gauge choice.
\newpage
\begin{flushleft}{\large{ \bf Acknowledgements}}
\end{flushleft}
We are grateful to Tim Jones for helpful conversations. 
\vskip 10pt
\appendix
\section{Details of the calculation}
The terms in Eqs.~\eqref{Hdef} which are independent of the choice Eq.~\eqref{chjj}, \eqref{chjja}, or Eq.~\eqref{Pforma:main}, are given by
\be
\Dhat=\left(\begin{matrix}0&0\cr
0&-\xi R+\tfrac{1}{2}\phi^2\lambda\end{matrix}\right),
\ee
and
\be
\Vhat^{\alpha\beta}=\left(\begin{matrix}\Vhat^{\alpha\beta}_{\hbar\hbar}&\Vhat^{\alpha\beta}_{\hbar h}\cr
                                                                    \Vhat^{\alpha\beta}_{h\hbar}&\Vhat^{\alpha\beta}_{hh}\end{matrix}\right),\quad
  \Uhat=\left(\begin{matrix}\Uhat_{\hbar\hbar}&\Uhat_{\hbar h}\cr
                                                                   \ Uhat_{h\hbar}&\Uhat_{hh}\end{matrix}\right),
\ee
where
\begin{subequations}\label{UV:main}
\begin{align}
\Vhat^{\alpha\beta,\mu\nu,\rho\sigma}_{\hbar\hbar}=&4g^{\alpha\beta}R^{\mu\rho\nu\sigma}+2\delta^{\mu\nu,\rho\sigma}(R^{\alpha\beta}-\tfrac13(1+\omega)Rg^{\alpha\beta})\nn
&-4g^{\mu\rho}(R^{\nu(\alpha}g^{\beta)\sigma}+R^{\sigma(\alpha}g^{\beta)\nu})\nn
&+\tfrac43(1+\omega)(R^{\rho\sigma}\delta^{\mu\nu,\alpha\beta}+R^{\mu\nu}\delta^{\rho\sigma,\alpha\beta}
+Rg^{\mu\rho}\delta^{\alpha\beta,\nu\sigma})\nn
&+\mu\xi\phi^2\left(\tfrac12g^{\alpha\beta}\delta^{\mu\nu,\rho\sigma}-g^{\mu\rho}\delta^{\alpha\beta,\nu\sigma}\right),\\
\Vhat^{\alpha\beta,\rho\sigma}_{\hbar h}=&\frac{1}{\sqrt{-\beta}}(-\omega R^{\rho\sigma}g^{\alpha\beta}
+\tfrac14\mu\xi\phi^2\delta^{\rho\sigma,\alpha\beta}),\\
\Vhat^{\alpha\beta,\mu\nu}_{h\hbar}=&\frac{1}{\sqrt{-\beta}}(-\omega R^{\mu\nu}g^{\alpha\beta}+\tfrac14\mu\xi\phi^2\delta^{\mu\nu,\alpha\beta}),\\
\Vhat^{\alpha\beta}_{hh}=&\frac{1}{4}g^{\alpha\beta}\left(\frac{\omega}{\beta} R+\tfrac34\mu\xi\phi^2\right),\\
U_{\hbar\hbar}=&2\mu\Bigl(-\tfrac14\delta^{\mu\nu,\rho\sigma}\nab^{\alpha}\phi\nab_{\alpha}\phi+g^{\mu\rho}\nab^{\nu}\phi\nab^{\sigma}\phi\nn
&-\tfrac14\xi R\phi^2\delta^{\mu\nu,\rho\sigma}+\xi R^{\mu\sigma}g^{\nu\rho}\phi^2\nn
&+\tfrac14\xi(R^{\mu\rho\nu\sigma}-R^{\mu\sigma}g^{\nu\rho})\phi^2\nn
&+\tfrac{1}{48}\delta^{\mu\nu,\rho\sigma}\lambda\phi^4\Bigr)+\ldots,\label{UV:a}\\
U_{hh}=&2\mu\left(\frac{1}{192\beta}\lambda\phi^4-\frac{1}{8\beta}\xi R\phi^2\right)+\ldots\label{UV:b}
\end{align}
\end{subequations}
The ellipses in Eqs.~\eqref{UV:a}, \eqref{UV:b} denote $\phi$-independent terms which play no r\^ole in our current calculation.
We draw the reader's attention to the terms on the 3rd line of our expression for $U_{\hbar\hbar}$ in Eq.~\eqref{UV:a}, which are not present in the corresponding expression in Eq.~(9.7) of Ref.~\cite{BOS}. They arise since the expansion of $R$ in (8.82) of Ref.~\cite{BOS} produces a term $-\tfrac12h^{\mu\nu}\nab_{\lambda}\nab_{\mu}h_{\lambda\nu}$. This represents a contribution to 
$\Vhat_{\hbar\hbar}$. However, this is assumed in the rest of the calculation to be symmetrised; and the symmetrisation generates the extra terms in $\Uhat_{\hbar\hbar}$.

In the case of the choice Eq.~\eqref{Pforma:main}, we have in Eq.~\eqref{Pform:main}
\begin{subequations}\label{pqbos:main}
\begin{align}
\phat_1^{\alpha\beta}=\qhat_1^{T\alpha\beta}=\Phat_1^{\alpha\beta}=&i\xi\phi\sqrt{2\mu}\left(\begin{matrix} 0&0\cr
                                                                         -\delta^{\mu\nu,\alpha\beta}&\frac{3}{4\sqrt{-\beta}}g^{\alpha\beta}
\end{matrix}\right),\label{pqbos:a}\\
\qhat_2^{\alpha}=-\phat_2^{T\alpha}=\Qhat_2^{\alpha}=&i\sqrt{2\mu}\left(\begin{matrix} 0&
(1-\xi)g^{\alpha\nu}\nab^{\mu}\phi\cr 0&\frac{1}{4\sqrt{-\beta}}(3\xi-1)\nab^{\alpha}\phi
\end{matrix}\right),\label{pqbos:a}\\
\phat_3=\Phat_3=&i\sqrt{2\mu}\left(\begin{matrix}0&0\cr
-\nab^{\mu}\nab^{\nu}\phi+\xi R^{\mu\nu}\phi
& \frac{1}{4\sqrt{-\beta}}(\nab^2\phi-\xi R\phi)\cr
&+\frac{1}{12\sqrt{-\beta}}\lambda\phi^3
\end{matrix}\right),\label{pqbos:b}\\
\qhat_3=\Qhat_3=&i\sqrt{2\mu}\left(\begin{matrix}0&\xi R^{\rho\sigma}\phi\cr
0& -\frac{1}{4\sqrt{-\beta}}\xi R\phi
+\frac{1}{12\sqrt{-\beta}}\lambda\phi^3
\end{matrix}\right),\\
\end{align}
\end{subequations} 
while in the case of the choice Eq.~\eqref{chjj}, \eqref{chjja}, we have in Eq.~\eqref{Pform:main}
\begin{subequations}\label{pqjj:main}
\begin{align}
\phat_1^{\alpha\beta}=\Pcal_1^{\alpha\beta}=2\Phat_1^{\alpha\beta}=&2i\xi\phi\sqrt{2\mu}\left(\begin{matrix} 0&0\cr
                                                                         -\delta^{\mu\nu,\alpha\beta}&\frac{3}{4\sqrt{-\beta}}g^{\alpha\beta}
\end{matrix}\right),\label{pqjj:a}\\
\qhat_1^{\alpha\beta}=&\phat_2^{\alpha}=0,\\
\qhat_2^{\alpha}=\Qcal_2^{\alpha}=&i\sqrt{2\mu}\left(\begin{matrix} 0&g^{\alpha\rho}\nab^{\sigma}\phi\cr
0&-\frac{1}{4\sqrt{-\beta}}\nab^{\alpha}\phi
\end{matrix}\right),\label{pqjj:a}\\ 
\phat_3=\Pcal_3=&i\sqrt{2\mu}\left(\begin{matrix}0&0\cr
\xi R^{\mu\nu}\phi
& -\frac{1}{4\sqrt{-\beta}}\xi R\phi\cr
&+\frac{1}{12\sqrt{-\beta}}\lambda\phi^3
\end{matrix}\right),\label{pqjj:b}\\
\qhat_3=\Qcal_3=&\Qhat_3.
\end{align}
\end{subequations}
Notice that $\Qcal_2^{\alpha}$ in Eq.~\eqref{pqjj:a} is as $\Qhat_2^{\alpha}$ in Eq.~\eqref{pqbos:a}, but without the terms involving $\xi$; and $\Pcal_3$ in Eq.~\eqref{pqjj:b} is the same as $\Phat_3$ in Eq.~\eqref{pqbos:b}, but without the terms with two $\phi$ derivatives. Finally, $\Qcal_3$ and $\Qhat_3$ are exactly the same.

As explained before, the procedure is to substitute Eq.~\eqref{Pbox} into the various terms of Eq.~\eqref{exp} and then use Eq.~\eqref{comm} to commute all factors of $\frac{1}{\Box}$ to the right-hand side. We are then able to apply the following ``universal functional traces'' tabulated in Ref.~\cite{BV}. 
\begin{subequations}
\begin{align}
\frac{\onehat}{\Box}\delta(x,y)|^{\rm{div}}_{y=x}=&\frac{2i}{16\pi^2\epsilon}\sqrt{-g}(\tfrac16R\onehat),\nn
\nab_{\mu_1}\nab_{\mu_2}\ldots\nab_{\mu_{2n-4}}\frac{\onehat}{\Box^n}\delta(x,y)|^{\rm{div}}_{y=x}=&-\frac{2i}{16\pi^2\epsilon}\sqrt{-g}\frac{\onehat g^{(n-2)}_{\mu_1\mu_2\ldots \mu_{2n-4}}}{2^{n-2}(n-1)!},\\
\nab_{\mu}\nab_{\nu}\frac{\onehat}{\Box^2}\delta(x,y)|^{\rm{div}}_{y=x}=&-\frac{2i}{16\pi^2\epsilon}\sqrt{-g}
\left[\tfrac16(R_{\mu\nu}-\tfrac12Rg_{\mu\nu})\onehat+\tfrac12\Rcal_{\mu\nu}\right],\\
\nab_{\mu}\nab_{\nu}\nab_{\rho}\nab_{\sigma}\frac{\onehat}{\Box^3}\delta(x,y)|^{\rm{div}}_{y=x}=&-\frac{2i}{16\pi^2\epsilon}\sqrt{-g}\tfrac14\Bigl\{-\tfrac23R_{(\nu\rho\sigma)\mu}\onehat+g_{\mu\nu}\left(\tfrac16R_{\rho\sigma}\onehat+\tfrac12\Rcal_{\rho\sigma}\right)\nn
&+g_{\nu\sigma}\left(\tfrac16R_{\mu\rho}\onehat+\tfrac12\Rcal_{\mu\rho}\right)+g_{\rho\nu}\left(\tfrac16R_{\mu\sigma}\onehat+\tfrac12\Rcal_{\mu\sigma}\right)\nn
&+g_{\sigma\mu}\left(\tfrac16R_{\nu\rho}\onehat+\tfrac12\Rcal_{\nu\rho}\right)+g_{\rho\mu}\left(\tfrac16R_{\nu\sigma}\onehat+\tfrac12\Rcal_{\nu\sigma}\right)\nn
&+g_{\rho\sigma}\left(\tfrac16R_{\mu\nu}\onehat+\tfrac12\Rcal_{\mu\nu}\right)-\tfrac{1}{12}Rg^{(2)}_{\mu\nu\rho\sigma}\onehat\Bigr\},
\label{nabid}
\end{align}
\end{subequations}
where
\be
g^{(n)}_{\mu_1\ldots\mu_{2n}}=\frac{(2n)!}{2^nn!}g_{(\mu_1\mu_2}g_{\mu_3\mu_4}\ldots g_{\mu_{2n-1}\mu_{2n})},
\ee
and 
\be
(\R_{\alpha\beta})_{\mu\nu}{}^{\rho\sigma}=2R_{\alpha\beta(\mu}{}^{\rho}\delta_{\nu)}{}^{\sigma}.
\label{Rcal}
\ee

We give here the results which do not depend whether the choice Eqs.~\eqref{chjj}, \eqref{chjja}, or Eq.~\eqref{Pforma:main} is made for $\phat_1^{\alpha\beta}$, $\qhat_1^{\alpha\beta}$, $\phat_2^{\alpha}$, $\qhat_2^{\alpha}$ in Eq.~\eqref{Pform:main}.  This of course includes cases where the term in Eq.~\eqref{exp} does not contain these quantities; but we also note that whichever choice is made, we have up to integration by parts
\begin{subequations}\label{sim:main}
\begin{align}
\phat_1\equiv&\Phat_1^{\alpha\beta}\nab_{\alpha}\nab_{\beta} +\hbox{one or zero derivative terms},\\
\qhat_1\equiv&\Phat_1^{T\alpha\beta}\nab_{\alpha}\nab_{\beta} +\hbox{one or zero derivative terms},
\end{align}
\end{subequations}
where $\Phat_1^{\alpha\beta}$ is defined in Eq.~\eqref{pqbos:a}. The terms with fewer derivatives will be finite if there are fewer than two derivatives altogether.
 The results are the following.
\begin{subequations}\label{noPQ:main}
\begin{align}
\Tr U|_{\rm{div}}=&\frac{2i}{16\pi^2\epsilon}\Tr(-\Uhat)\nn
 =&\frac{2i}{16\pi^2\epsilon}\left\{-\mu\left(9+\frac{1}{4\beta}\right)\tfrac{1}{24}\lambda\phi^4+3\mu\xi\tfrac12R\phi^2\right\},\label{noPQ:x}\\ 
\Tr V|_{\rm{div}} =&\frac{2i}{16\pi^2\epsilon}\Tr\left(\tfrac{1}{12}R\Vhat^{\alpha}{}_{\alpha}-\tfrac16R_{\alpha\beta}\Vhat^{\alpha\beta}\right)\nn
=&\frac{2i}{16\pi^2\epsilon}\mu\xi\left(\frac34+\frac{1}{16\beta}\right)\tfrac12R\phi^2,\label{noPQ:b}\\
-\tfrac12\Tr V^2|_{\rm{div}}=&\frac{2i}{16\pi^2\epsilon}\Tr\left(\tfrac{1}{24}\Vhat_{\alpha\beta}\Vhat^{\alpha\beta}
+\tfrac{1}{48}\Vhat^{\alpha}{}_{\alpha}\Vhat^{\beta}{}_{\beta}\right)\nn
=&\frac{2i}{16\pi^2\epsilon}\Bigl\{\mu^2\xi^2\left(\frac{63}{4}-\frac{9}{8\beta}+\frac{27}{64\beta^2}\right)\tfrac{1}{24}\phi^4\nn
&+\mu\xi\left(-\frac72\omega-2+\frac{3\omega}{32\beta^2}\right)\tfrac12R\phi^2\Bigr\},\label{noPQ:c}\\
\Tr\left(-\tfrac12D^2+D\right)|_{\rm{div}}=&\frac{2i}{16\pi^2\epsilon}\left(3\tfrac{1}{24}\lambda^2\phi^4
-(\xi-\tfrac16)\lambda\tfrac12R\phi^2\right),\nn
\Tr Vq_1p_1|_{\rm{div}}=&\frac{2i}{16\pi^2\epsilon}\Tr\left(-\tfrac{1}{192}\Vhat^{\alpha\beta}P_1^{T\mu\nu}P_1^{\rho\sigma}g^{(3)}_{\alpha\beta\mu\nu\rho\sigma}\right)\nn
=&\frac{2i}{16\pi^2\epsilon}\Bigl\{\mu\xi^2\left(-3+\omega-\frac{9\omega}{16\beta^2}\right)\tfrac12
R\phi^2\nn
&+\mu^2\xi^3\left(-\frac38+\frac{9}{16\beta}-\frac{27}{128\beta^2}\right)\phi^4\Bigr\},\label{noPQ:d}\\
\Tr q_1Dp_1|_{\rm{div}}=&\frac{2i}{16\pi^2\epsilon}\left(-\tfrac{1}{24}\Phat_1^{T\mu\nu}\Dhat\Phat_1^{\rho\sigma}g^{(2)}_{\mu\nu\rho\sigma}\right)\nn
=&\frac{2i}{16\pi^2\epsilon}\Bigl\{\mu\xi^3\left(-3+\frac{9}{4\beta}\right)\tfrac12R\phi^2\nn
&+\mu\xi^2\left(18-\frac{27}{2\beta}\right)\tfrac{1}{24}\lambda\phi^4\Bigr\},\\
-\tfrac12\Tr q_1p_1q_1p_1|_{\rm{div}}=&\frac{2i}{16\pi^2\epsilon}\Tr\tfrac{1}{3840}\Phat_1^{T\alpha\beta}\Phat_1^{\gamma\delta}\Phat_1^{T\mu\nu}\Phat_1^{\rho\sigma}g^{(4)}_{\alpha\beta\gamma\delta\mu\nu\rho\sigma}\nn
=&\frac{2i}{16\pi^2\epsilon}\mu^2\xi^4\left(\frac98-\frac{27}{16\beta}+\frac{81}{128\beta^2}\right)\phi^4,\label{noPQ:y}
\end{align}
\end{subequations}
For the terms involving $p_1$ or $q_1$, the divergences may be obtained by substituting Eq.~\eqref{sim:main} and simply retaining the terms given explicitly on the right-hand side; and as mentioned earlier, are the same whichever choice is made in Eq.~\eqref{Pform:main}.

The corresponding results in Ref.~\cite{BOS} contain several minor errors, mostly typographical. We have inserted a $\tfrac12$ before the $R\phi^2$ in Eqs.~\eqref{noPQ:b}, \eqref{noPQ:c}; and changed $\beta$ to $\beta^2$ in the third term of Eq.~\eqref{noPQ:d}.

The most complex evaluation is that of $\Tr(p_1q_1)$. This differs significantly depending on the choice of Eq.~\eqref{chjj} or Eq.~\eqref{Pforma:main}. However both choices involve a basic nucleus given by
\be
T=-\Tr\Bigl(\Phat_1^{T\gamma\delta}\nab_{\gamma}\nab_{\delta}\frac{1}{\Box}\Phat_1^{\alpha\beta}\nab_{\alpha}\nab_{\beta}\frac{1}{\Box^2}\Bigr),
\ee 
where $\Phat_1^{\alpha\beta}$ is given in Eqs.~\eqref{pqjj:a} or Eq.~\eqref{pqbos:a}. We shall accordingly describe the treatment of this common term in some detail here, before describing the differences in the two approaches in the respective Appendices.
Using
\be
\frac{1}{\Box}\Phat^{\alpha\beta}_1\nab_{\alpha}\nab_{\beta}
=\Phat^{\alpha\beta}_1\nab_{\alpha}\nab_{\beta}\frac{1}{\Box}
-\frac{1}{\Box}[\Box,\Phat^{\alpha\beta}_1\nab_{\alpha}\nab_{\beta}]\frac{1}{\Box}
\ee
where\footnote{Since $\frac{1}{\Box}$ is a shorthand for the Green function $G(x,x')$, derivatives on opposite sides are defined at different points. A more rigorous but lengthier treatment requires starting from $\nab_{\alpha}\nab_{\beta}=\int dv_{x'}\nab_{\alpha'}\nab_{\beta'}\Box'G(x',x)g^{(\alpha'}{}_{\alpha}g^{\beta')}{}_{\beta}$, where $g_{\mu\nu'}$ is the bivector giving parallel transport of vectors along the geodesic from $x'$ to $x$. Using properties of $g_{\mu\nu'}$ as given in Ref.~\cite{DeWitt}, we have checked that the same result is obtained.
}
\begin{align}
[\Box,\Phat^{\alpha\beta}_1\nab_{\alpha}\nab_{\beta}]\onehat=&(\Box \Phat^{\alpha\beta}_1)\nab_{\alpha}\nab_{\beta}\onehat
+2(\nab^{\kappa}\Phat^{\alpha\beta}_1)\nab_{\kappa}\nab_{\alpha}\nab_{\beta}\onehat\nn
&+2\Phat_1^{\alpha\beta}(R_{\alpha}{}^{\kappa}\nab_{\beta}\nab_{\kappa}-R_{\alpha}{}^{\kappa}{}_{\beta}{}^{\lambda}\nab_{\kappa}\nab_{\lambda}-2\R_{\alpha}{}^{\kappa}\nab_{\beta}\nab_{\kappa})\onehat+\ldots
\label{excomm}
\end{align}
and similarly
\be
\frac{1}{\Box}(\nab^{\kappa}\Phat^{\alpha\beta}_1)\nab_{\kappa}\nab_{\alpha}\nab_{\beta}
=(\nab^{\kappa}\Phat^{\alpha\beta}_1)\nab_{\kappa}\nab_{\alpha}\nab_{\beta}\frac{1}{\Box}
-2\frac{1}{\Box}(\nab^{\kappa}\nab^{\lambda}\Phat^{\alpha\beta}_1)\nab_{\kappa}\nab_{\lambda}\nab_{\alpha}\nab_{\beta}\frac{1}{\Box}+\ldots
\label{excomma}
\ee
we find 
\begin{align}
T=&-\Tr\Bigl(\Phat_1^{T\gamma\delta}\Bigl[\Phat^{\alpha\beta}_1\nab_{\gamma}\nab_{\delta}\nab_{\alpha}\nab_{\beta}\frac{\onehat}{\Box^3}+(\nab_{\gamma}\nab_{\delta}\Phat^{\alpha\beta}_1)\nab_{\alpha}\nab_{\beta}\frac{\onehat}{\Box^3}\nn
&+4(\nab^{\kappa}\nab^{\lambda}\Phat^{\alpha\beta}_1)\nab_{\kappa}\nab_{\lambda}\nab_{\alpha}\nab_{\beta}\nab_{\gamma}\nab_{\delta}\frac{\onehat}{\Box^5}\nn&-4(\nab_{\gamma}\nab^{\kappa}\Phat^{\alpha\beta}_1)\nab_{\delta}\nab_{\kappa}\nab_{\alpha}\nab_{\beta}\frac{\onehat}{\Box^4}
-\Box\Phat^{\alpha\beta}_1\nab_{\alpha}\nab_{\beta}\nab_{\gamma}\nab_{\delta}\frac{\onehat}{\Box^4}\nn
&+2\Phat^{\alpha\beta}_1(R_{\alpha}{}^{\kappa}{}_{\beta}{}^{\lambda}\nab_{\kappa}\nab_{\lambda}-R_{\alpha}{}^{\kappa}\nab_{\beta}\nab_{\kappa}+2\R_{\alpha}{}^{\kappa}\nab_{\beta}\nab_{\kappa})\nab_{\gamma}\nab_{\delta}\frac{\onehat}{\Box^4}\Bigr]\Bigr),
\end{align}
so that
\begin{align}
T|_{\rm{div}}=&\frac{2i}{16\pi^2\epsilon}\Tr\Bigl\{\Phat_1^{T\gamma\delta}\Bigl(\tfrac18\nab_{\gamma}\nab_{\delta}\Phat_1^{\alpha}{}_{\alpha}-\tfrac{1}{48}g_{\gamma\delta}\Box\Phat^{\alpha}_1{}_{\alpha}\nn
&-\tfrac{1}{24}\Box\Phat_{1\gamma\delta}+\tfrac{1}{24}g_{\gamma\delta}\nab_{\alpha}\nab_{\beta}\Phat_1^{\alpha\beta}-\tfrac16\nab_{\alpha}\nab_{\gamma}\Phat^{\alpha}_1{}_{\delta}\nn
&+\tfrac{1}{24}R_{\gamma\delta}\Phat^{\alpha}_{1\alpha}+\tfrac{1}{24}g_{\gamma\delta}R_{\alpha\beta}\Phat_1^{\alpha\beta}-\tfrac{1}{48}g_{\gamma\delta}R\Phat_1^{\alpha}{}_{\alpha}\nn
&-\tfrac{1}{24}R\Phat_{1\gamma\delta}-\tfrac16R_{\alpha\gamma\beta\delta}\Phat_1^{\alpha\beta}+\tfrac16R_{\gamma\alpha}\Phat_1^{\alpha}{}_{\delta}\Bigr)\Bigr\}\nn
&+\frac{2i}{16\pi^2\epsilon}\Tr\Bigl(\frac{1}{6\epsilon}\Phat_1^{T\alpha\gamma}\Phat_1^{\beta}{}_{\gamma}\R_{\alpha\beta}
+\frac{1}{6\epsilon}(R_{\alpha\gamma\beta\delta}
\Phat_1^{T\alpha\beta}\Phat_1^{\gamma\delta}-\Phat_1^{T\alpha\gamma}\Phat_1^{\beta}{}_{\gamma}R_{\alpha\beta})\Bigr)\nn
=&\frac{2i}{16\pi^2\epsilon}\mu\xi^2\Bigl\{-\tfrac32g^{\mu\nu}\pa_{\mu}\phi\pa_{\nu}\phi-\left(\frac12+\frac{3}{8\beta}\right)\tfrac12R\phi^2+(2+2)\tfrac12R\phi^2\Bigr\}.
\label{Tres}
\end{align}
In Eqs.~\eqref{excomm}, \eqref{excomma}, the ellipses indicated terms involving two Riemann tensors, derivatives of Riemann tensors, or derivatives of $P^{\alpha\beta}_1$ together with Riemann tensors, which do not contribute to divergences in our calculation. We have also used the symmetry of $P^{\alpha\beta}_1$ in $\alpha$, $\beta$.
As emphasised before, 
we have included in Eq.~\eqref{excomm} terms resulting from commuting $\Box$ with the derivatives $\nab_{\alpha,\beta}$; and in doing this we have used
\be
[\nab_{\alpha},\nab_{\beta}]\onehat=\R_{\alpha\beta}\onehat.
\ee
We have also included the contributions deriving from the terms involving $\R$ in Eq.~\eqref{nabid}, which were overlooked in Ref.~\cite{BOS}. In Eq.~\eqref{Tres} we have explicitly displayed these extra $R\phi^2$ terms which we have identified.

\section{The original calculation corrected}
In this Appendix we follow the approach of Ref.~\cite{BOS} in symmetrising $\phat_1^{\alpha\beta}$, $\phat_1^{\alpha\beta}$, $\phat_2^{\alpha}$, $\qhat_2^{\alpha}$ according to Eq.~\eqref{Pforma:main}, and hence taking these quantities as in Eq.~\eqref{pqbos:main}.
However we correct the calculations of Ref.~\cite{BOS} by correctly distinguishing between left and right acting derivatives in the definitions of $\phat_1$, $\qhat_1$, $\phat_2$, $\qhat_2$, as encapsulated in Eq.~\eqref{Pform:main}.
Eq.~\eqref{comm} may be rewritten as 
\be
\frac{1}{\Box}X
=X\frac{1}{\Box}-\frac{1}{\Box}[(\Box X)+2(\nab^{\alpha}X)\nab_{\alpha}]\frac{1}{\Box}
\ee
which may be used to derive the useful results
\begin{subequations}\label{commid:main}
\begin{align}
\Tr X[\nab_{\alpha}\frac{1}{\Box}Y\nab_{\beta}\nab_{\gamma}\frac{1}{\Box^2}]
=&-\frac{2i}{16\pi^2\epsilon}\frac{1}{12}\Tr [X(2\delta_{\beta\gamma}\nab_{\alpha}Y-\delta_{\alpha\beta}\nab_{\gamma}Y
-\delta_{\alpha\gamma}\nab_{\beta}Y)],\\
\Tr X[\nab_{\alpha}\frac{1}{\Box^2}Y\nab_{\beta}\nab_{\gamma}\frac{1}{\Box}]
=&-\frac{2i}{16\pi^2\epsilon}\frac{1}{12}\Tr [X(\delta_{\beta\gamma}\nab_{\alpha}Y-2\delta_{\alpha\beta}\nab_{\gamma}Y
-2\delta_{\alpha\gamma}\nab_{\beta}Y)].
\end{align}
\end{subequations}
Using  \eqref{commid:main}, we obtain
\begin{subequations}
\begin{align}
-\Tr( p_1q_1)=&-\tfrac14\Tr\left[(\Phat_1^{\alpha\beta}\nab_{\alpha}\nab_{\beta}+\nabl_{\alpha}\nabl_{\beta}\Phat_1^{\alpha\beta})\frac{1}{\Box^2}
(\Phat_1^{T\gamma\delta}\nab_{\gamma}\nab_{\delta}+\nabl_{\gamma}\nabl_{\delta}\Phat_1^{T\gamma\delta})
\frac{1}{\Box}\right]\nn
=&T-\tfrac12\Tr\Bigl(\Phat_1^{\alpha\beta}\nab_{\alpha}\nab_{\beta}\frac{1}{\Box^2}(\nab_{\gamma}\nab_{\delta}\Phat_1^{T\gamma\delta})\frac{1}{\Box}
+(\nab_{\alpha}\nab_{\beta}\Phat_1^{\alpha\beta})\frac{1}{\Box^2}\Phat_1^{T\gamma\delta}\nab_{\gamma}\nab_{\delta}\frac{1}{\Box}
\nn
&+2\nab_{\beta}\Phat_1^{\alpha\beta}\nab_{\alpha}\frac{1}{\Box^2}\Phat_1^{T\gamma\delta}\nab_{\gamma}\nab_{\delta}\frac{1}{\Box}
+2\Phat_1^{\alpha\beta}\nab_{\alpha}\nab_{\beta}\frac{1}{\Box^2}(\nab_{\delta}\Phat_1^{T\gamma\delta)}\nab_{\gamma}\frac{1}{\Box}\nn
&+2(\nab_{\beta}\Phat_1^{\alpha\beta})\nab_{\alpha}\frac{1}{\Box^2}(\nab_{\delta}\Phat_1^{T\gamma\delta)}\nab_{\gamma}\frac{1}{\Box}\Bigr),\\
-\Tr(p_1q_2+p_2q_1)=&-\tfrac14\Tr\Bigl[(\Phat_1^{\gamma\delta}\nab_{\gamma}\nab_{\delta}+\nabl_{\gamma}\nabl_{\delta}\Phat_1^{\gamma\delta})\frac{1}{\Box^2}(\Qhat_2^{\alpha}\nab_{\alpha}-\nabl_{\alpha}\Qhat_2^{\alpha})\frac{1}{\Box}\nn
&-(\Qhat_2^{T\alpha}\nab_{\alpha}-\nabl_{\alpha}\Qhat_2^{T\alpha})\frac{1}{\Box^2}(\Phat_1^{T\gamma\delta}\nab_{\gamma}\nab_{\delta}+\nabl_{\gamma}\nabl_{\delta}\Phat_1^{T\gamma\delta})\frac{1}{\Box}\Bigr]\nn
=&-\Tr\Bigl[\Phat_1^{\gamma\delta}\nab_{\gamma}\nab_{\delta}\frac{1}{\Box^2}\Qhat_2^{\alpha}\nab_{\alpha}\frac{1}{\Box}-\Qhat_2^{T\alpha}\nab_{\alpha}\frac{1}{\Box^2}\Phat_1^{T\gamma\delta}\nab_{\gamma}\nab_{\delta}\frac{1}{\Box}\Bigr]\nn
&-\tfrac12\Tr\Bigl[\Phat_1^{\gamma\delta}\nab_{\gamma}\nab_{\delta}\frac{1}{\Box^2}(\nab_{\alpha}\Qhat_2^{\alpha})\frac{1}{\Box}-(\nab_{\alpha}\Qhat_2^{T\alpha})\frac{1}{\Box^2}\Phat_1^{T\gamma\delta}\nab_{\gamma}\nab_{\delta}\frac{1}{\Box}\nn
&+2(\nab_{\delta}\Phat_1^{\gamma\delta})\nab_{\gamma}\frac{1}{\Box^2}\Qhat_2^{\alpha}\nab_{\alpha}\frac{1}{\Box}-2\Qhat_2^{T\alpha}\nab_{\alpha}\frac{1}{\Box^2}(\nab_{\delta}\Phat_1^{T\gamma\delta})\nab_{\gamma}\frac{1}{\Box}\Bigr],
\end{align}
\end{subequations}
so that using Eqs.~\eqref{Pform:main}, \eqref{Pforma:main} we find
\begin{subequations}\label{BOSres:main}
\begin{align}
-\Tr( p_1q_1)|_{\rm{div}}=&T|_{\rm{div}}+\frac{2i}{16\pi^2\epsilon}\Tr\Bigl\{\frac{1}{8}[\Phat_1^{\alpha}{}_{\alpha}(\nab_{\gamma}\nab_{\delta}\Phat_1^{T\gamma\delta})+(\nab_{\alpha}\nab_{\beta}\Phat_1^{\alpha\beta})\Phat_1^{T\gamma}{}_{\gamma}\nn
&+2(\nab_{\beta}\Phat_1^{\alpha\beta})(\nab_{\delta}\Phat_1^{T\delta}{}_{\alpha})]+\frac{1}{12}\nab_{\beta}\Phat_1^{\alpha\beta}(\nab_{\alpha}\Phat_1^{T\gamma}{}_{\gamma}
-4\nab^{\gamma}\Phat^T_{1\alpha\gamma})\nn
&+\frac16\nab_{\delta}\Phat_1^{T\beta\delta}(\nab_{\beta}\Phat_1^{\alpha}{}_{\alpha}-\nab_{\alpha}\Phat_1^{\alpha\beta})\Bigr\}\nn
=&T|_{\rm{div}}+\frac{2i}{16\pi^2\epsilon}\tfrac14\Phat^{T\rho\sigma}_1\nab_{\mu}\nab_{\rho}\Phat^{\mu}_1{}_{\sigma}\nn
=&\frac{2i}{16\pi^2\epsilon}\mu\xi^2\Bigl\{-\tfrac32g^{\mu\nu}\pa_{\mu}\phi\pa_{\nu}\phi-\left(\frac12+\frac{3}{8\beta}\right)\tfrac12R\phi^2\nn
&+\left(\frac94-\frac{9}{16\beta}\right)\tfrac12g^{\mu\nu}\pa_{\mu}\phi\pa_{\nu}\phi+(6-2)\tfrac12R\phi^2\Bigr\}, \label{BOSres:x}\\
-\Tr(p_1q_2+p_2q_1)|_{\rm{div}}=&\frac{2i}{16\pi^2\epsilon}\Tr \Bigl[\tfrac16(\nab_{\alpha}\Phat_1^{\beta}{}_{\beta}\Qhat_2^{\alpha}-\nab_{\beta}\Phat_1^{\alpha\beta}\Qhat_{2\alpha})\nn
&-\tfrac{1}{12}(\Qhat_2^{T\alpha}\nab_{\alpha}\Phat_1^{T\beta}{}_{\beta}-4\Qhat^T_{2\alpha}\nab_{\beta}\Phat_1^{T\alpha\beta})\Bigr]\nn
&+\frac{2i}{16\pi^2\epsilon}\tfrac18\Tr\Bigl[P_1^{\gamma}{}_{\gamma}\nab_{\alpha}Q_2^{\alpha}
-\nab_{\alpha}Q_2^{T\alpha}P_1^{T\gamma}{}_{\gamma}\nn
&+2\nab_{\gamma}P_1^{\gamma}{}_{\alpha}Q_2^{\alpha}-2Q_2^{T\alpha}\nab_{\gamma}P_1^{T\gamma}{}_{\alpha}\Bigr]\nn
=&\frac{2i}{16\pi^2\epsilon}\left\{3\mu\xi\left(\frac{1-\xi}{2}+\frac{3\xi-1}{8\beta}\right)\tfrac12g^{\mu\nu}\pa_{\mu}\phi\pa_{\nu}\phi\right\},\label{BOSres:z}\\
-\Tr (p_2q_2)|_{\rm{div}}=&\frac{2i}{16\pi^2\epsilon}\Tr\left(-\tfrac14\Qhat_2^{T\alpha}\Qhat_{2\alpha}\right)\nn
=&\frac{2i}{16\pi^2\epsilon}\mu\left\{\frac{9}{4}(\xi-1)^2-\frac{1}{16\beta}(3\xi-1)^2\right\}\tfrac12g^{\mu\nu}\pa_{\mu}\phi\pa_{\nu}\phi,\\
-\Tr (q_3p_1+q_1p_3)|_{\rm{div}}=&\frac{2i}{16\pi^2\epsilon}\Tr\tfrac14\left(\Qhat_3\Phat_1^{\alpha}{}_{\alpha}+\Phat_1^{T\alpha}{}_{\alpha}\Phat_3\right)\nn
=&\frac{2i}{16\pi^2\epsilon}\mu\Bigl\{-\frac{3\xi}{4\beta}\tfrac12g^{\mu\nu}\pa_{\mu}\phi\pa_{\nu}\phi-\frac{3\xi^2}{2\beta}\tfrac12R\phi^2+\frac{6\xi}{\beta}\tfrac{1}{24}\lambda\phi^4\Bigr\},
\label{BOSres:y}
\end{align}
\end{subequations}
where we have used Eq.~\eqref{nabid} throughout, and Eq.~\eqref{commid:main} in the derivation of Eqs.~\eqref{BOSres:x} and \eqref{BOSres:z}. We have corrected the numerical factor from Ref.~\cite{BOS} in the last term of Eq.~\eqref{BOSres:y}. In Eq.~\eqref{BOSres:x} we have explicitly displayed the extra $R\phi^2$ terms which we identified in Eq.~\eqref{Tres}.

The effect of our changes on the $g^{\mu\nu}\pa_{\mu}\phi\pa_{\nu}\phi$ terms has been as follows. The terms of this form in $\Tr(p_2q_2)$, $\Tr(q_2p_1+q_1p_2)$ and $\Tr(q_3p_1+q_1p_3)$ are unchanged from Ref.~\cite{BOS}; and the extra $\xi^2$ term in $\Tr(p_1q_1)$ is such as to cancel the entire original $\xi^2$ term from Ref.~\cite{BOS}. In sum, the $\xi^2$ terms have cancelled, leaving the $\xi$ and $\xi$-independent terms unchanged.

\section{The streamlined calculation}
In this section we apply the choice Eq.~\eqref{chjj} in Eq.~\eqref{Pform:main} and consequently use  Eq.~\eqref{pqjj:main}.
We have, after again using Eq.~\eqref{commid:main}, 
\begin{subequations}
\begin{align}
-\Tr p_1q_1=&-\Tr\left[\Phat_1^{\alpha\beta}\nab_{\alpha}\nab_{\beta}\frac{1}{\Box^2}\nabl_{\gamma}\nabl_{\delta}\Phat_1^{T\gamma\delta}\frac{1}{\Box}\right]\nn
=&-T-\Tr\Bigl(\Phat_1^{\alpha\beta}\nab_{\alpha}\nab_{\beta}\frac{1}{\Box^2}[\nab_{\gamma}\nab_{\delta}\Phat_1^{T\gamma\delta}]\frac{1}{\Box}\nn
&+2\Phat_1^{\alpha\beta}\nab_{\alpha}\nab_{\beta}\frac{1}{\Box^2}[\nab_{\gamma}\Phat_1^{T\gamma\delta}]\nab_{\delta}\frac{1}{\Box}\Bigr),\\
-\Tr(p_1q_2+p_2q_1)=&-\Tr\Bigl[\Phat_1^{\gamma\delta}\nab_{\gamma}\nab_{\delta}\frac{1}{\Box^2}\Qcal_2^{\alpha}\nab_{\alpha}\frac{1}{\Box}+\nabl_{\alpha}\Qcal_2^{T\alpha}\frac{1}{\Box^2}\nabl_{\gamma}\nabl_{\delta}\Phat_1^{T\gamma\delta}\frac{1}{\Box}\Bigr]\nn
=&-2\Tr\Bigl[\Phat_1^{\gamma\delta}\nab_{\gamma}\nab_{\delta}\frac{1}{\Box^2}\Qcal_2^{\alpha}\nab_{\alpha}\frac{1}{\Box}\Bigl],
\end{align}
\end{subequations}
so that 
\begin{subequations}\label{jjres:main}
\begin{align}
-\Tr (p_1q_1)|_{\rm{div}}=&-T|_{\rm{div}}+\frac{2i}{16\pi^2\epsilon}\Tr\Bigl(\frac{1}{4\epsilon}\Phat_1^{\alpha}{}_{\alpha}\nab_{\gamma}\nab_{\delta}\Phat_1^{T\gamma\delta}
+\frac{1}{3\epsilon}\nab_{\gamma}\Phat_1^{T\gamma\alpha}[\nab_{\alpha}\Phat_1^{\beta}{}_{\beta}
-\nab^{\beta}\Phat_{1\alpha\beta}]\Bigr)\nn
=&-T|_{\rm{div}}+\frac{2i}{16\pi^2\epsilon}\Tr\Bigl(-\frac{1}{12\epsilon}\Phat_1^{\alpha}{}_{\alpha}\nab_{\gamma}\nab_{\delta}\Phat_1^{T\gamma\delta}
+\frac{1}{3\epsilon}\nab^{\beta}\nab_{\gamma}\Phat_1^{T\gamma\alpha}
\Phat_{1\alpha\beta}\Bigr)\nn
=&
\frac{2i}{16\pi^2\epsilon}\mu\xi^2\Bigl\{-\tfrac32g^{\mu\nu}\pa_{\mu}\phi\pa_{\nu}\phi-\left(\frac12+\frac{3}{8\beta}\right)\tfrac12R\phi^2\nn
&+\tfrac32g^{\mu\nu}\pa_{\mu}\phi\pa_{\nu}\phi+4.\tfrac12R\phi^2\Bigr\},\nn
=&\frac{2i}{16\pi^2\epsilon}\mu\xi^2\left(\frac72-\frac{3}{8\beta}\right)\tfrac12R\phi^2,\label{jjres:x}\\
-\Tr(p_1q_2+p_2q_1)|_{\rm{div}}=&\frac{2i}{16\pi^2\epsilon}\tfrac13\Tr\Bigl(\nab_{\alpha}\Phat_1^{\beta}{}_{\beta}\Qcal_2^{\alpha}-\nab^{\beta}\Phat_{1\alpha\beta}\Qcal_2^{\alpha}\Bigr)\nn
=&-\frac{2i}{16\pi^2\epsilon}\mu\xi\left(3+\frac{3}{4\beta}\right)\tfrac12g^{\mu\nu}\pa_{\mu}\phi\pa_{\nu}\phi\\
-\Tr( p_2q_2)|_{\rm{div}}=&\frac{2i}{16\pi^2\epsilon}\Tr\left(-\tfrac14\Qcal^T_{2\alpha}\Qcal_2^{\alpha}\right)\nn
=&\frac{2i}{16\pi^2\epsilon}\mu\left\{\frac{9}{4}-\frac{1}{16\beta}\right\}\tfrac12g^{\mu\nu}\pa_{\mu}\phi\pa_{\nu}\phi,\\
-\Tr (q_3p_1+q_1p_3)|_{\rm{div}}=&\frac{2i}{16\pi^2\epsilon}\Tr\tfrac14\left(\Qhat_3\Phat_1^{\alpha}{}_{\alpha}+\Phat_1^{T\alpha}{}_{\alpha}\Phat_3\right)\nn
=&\frac{2i}{16\pi^2\epsilon}\mu\Bigl\{-\frac{3\xi^2}{2\beta}\tfrac12R\phi^2
+\frac{6\xi}{\beta}\tfrac{1}{24}\lambda\phi^4\Bigr\},\label{jjres:y}
\end{align}
\end{subequations}
where once again we have used Eqs.~\eqref{nabid} throughout, and Eq.~\eqref{commid:main} in the derivation of the 1st and 2nd results. Again, in Eq.~\eqref{jjres:x} we have explicitly displayed the extra $R\phi^2$ terms which we identified in Eq.~\eqref{Tres}.

The effect of our changes on the $g^{\mu\nu}\pa_{\mu}\phi\pa_{\nu}\phi$ terms has been as follows. There are now no contributions from $\Tr(p_1q_1)$. The terms proportional to $\xi^2$ and $\xi$ have disappeared from $\Tr(p_2q_2)$ and $\Tr(q_3p_1+q_1p_3)$, leaving the $\xi$-independent term (which comes only from $\Tr(p_2q_2)$) unchanged from Ref.~\cite{BOS}; and $\Tr(q_2p_1+q_1p_2)$  yields only a term proportional to $\xi$, which alone is now equal to the entire original term proportional to $\xi$ in Ref.~\cite{BOS}. In sum, the terms proportional to $\xi^2$ have cancelled, leaving the term proportional to $\xi$ and the $\xi$-independent term unchanged from Ref.~\cite{BOS}; as was found in Appendix B.

\end{document}